\documentclass[pra,twocolumn,showpacs,floatfix]{revtex4}
\usepackage{graphicx}
\begin{document}

\title{Solitary waves of Bose-Einstein condensed atoms
confined in finite rings}
\author{J. Smyrnakis$^1$, M. Magiropoulos$^1$, G. M. Kavoulakis$^1$,
and A. D. Jackson$^2$}
\affiliation{$^1$Technological Education Institute of Crete, P.O. 
Box 1939, GR-71004, Heraklion, Greece \\
$^2$The Niels Bohr International Academy, The Niels
Bohr Institute, Blegdamsvej 17, DK-2100, Copenhagen \O, Denmark}
\date{\today}

\begin{abstract}

Motivated by recent progress in trapping Bose-Einstein 
condensed atoms in toroidal potentials, we examine solitary-wave
solutions of the nonlinear Schr\"odinger equation subject to  
periodic boundary conditions. When the circumference of 
the ring is much larger than the size of the wave, the 
density profile is well approximated by that of an 
infinite ring, however the density and the velocity of 
propagation cannot vanish simultaneously. When the size 
of the ring becomes comparable to the size of the wave, 
the density variation becomes sinusoidal and the velocity 
of propagation saturates to a constant value.

\end{abstract}
 \maketitle

\section{Introduction}

The problem of solitary-wave solutions of the one-dimensional
nonlinear Schr\"odinger equation has been studied extensively, 
starting with the work of Zakharov and Shabat \cite{ZS} and of 
Tsuzuki \cite{Ts}. In addition, Lieb \cite{Lieb} has examined the 
problem of a Bose gas in strictly one dimension with periodic 
boundary conditions and a repulsive interaction, and he derived 
the excitation spectrum.  As demonstrated in that study, the 
spectrum consists of the usual Bogoliubov branch as well as a 
second branch. This branch was later identified as corresponding 
to solitary-wave excitation \cite{Kul,Ish}.

Bose-Einstein condensed atoms interacting via a contact potential
are described by a nonlinear Schr\"odinger equation and thus offer 
a very suitable system for the study of solitary waves \cite{Fra}. 
Solitary waves have been created in quasi-one-dimensional traps with 
a very weak harmonic potential in the long direction \cite{Han} and 
in pancake traps \cite{Phil}. Recently, it also became possible 
to confine atoms in toroidal traps \cite{Kurn,Olson,wvk,Hen,Phillips} 
in which persistent currents were observed \cite{Phillips}. 

Based on this experimental progress, it is natural to consider 
the possible creation of solitary waves in such toroidal traps.  
This is the subject of the present study.  In the following, 
we examine the properties of solitary waves, including their 
density, propagation velocity, and phase (and thus the superfluid 
velocity), with periodic boundary conditions. We note that 
Carr {\it et al.\ }\cite{Carr} have previously considered 
the special case of {\it static\/} periodic solitary-wave 
solutions for a fixed particle number and a length.

For simplicity we consider in the present study a very tight 
toroidal trap, neglecting the transverse degrees of freedom, and 
thus consider the purely one-dimensional motion of atoms on a ring. 
The main question to be considered here is the behavior of solitonic 
solutions as the length of the period $L$ is varied for fixed $N/L$, 
where $N$ is the number of particles. The relevant dimensionless 
parameter is the ratio of $L$ and the coherence length, $\xi$, 
which we choose as that corresponding to the maximum density.  
According to the value of $L/\xi$ we distinguish between ``large" 
rings, where $L/\xi \gg 1$, and ``small" rings, where $L/\xi \ll 1$. 
In Lieb's study \cite{Lieb}, the limit $N \to \infty$ and $L \to 
\infty$ with $N/L$ finite was considered. This corresponds to the 
limit of large rings in the present study. We note that the ratio 
$(\xi/L)^2$ also describes the ratio between the typical zero-point 
energy, $\hbar^2/(2 M L^2)$, and the typical interaction energy, 
$\hbar^2/(2 M \xi^2)$. Finally, we restrict ourselves to the case 
of repulsive effective interatomic interactions.

As we will show below, in the limit of large rings, the wave is 
localized within a length scale determined by the coherence length, 
and the density is constant over the rest of the system, in agreement 
with known results for an infinite system.  On the other hand, 
depending on the phase difference at the edges of the system 
(which is either zero or some integer multiple of $2 \pi$), two 
different situations can arise.  For zero phase difference, the 
minimum possible value of the density is zero, and the velocity of 
propagation varies between a lowest nonzero value, which scales as 
$\xi/L$ and essentially corresponds to the superfluid ``drift" 
velocity, and a maximum velocity which is set by the speed of 
sound. In the case of a nonzero phase difference, the velocity
of propagation can be zero, but the corresponding density cannot 
vanish, scaling like $(\xi/L)^2$.  Clearly, when $N \to \infty$ 
and $L \to \infty$ with $N/L$ finite, we find agreement with known 
results.

In the opposite limit of small rings, the density becomes sinusoidal 
with a length scale given by $L$ itself.  Furthermore, the 
velocity of propagation saturates to a constant value, which scales 
like $1/L$, and the minimum density can become zero. 

In what follows we first formulate our problem in Sec.\,II. 
Then, in Sec.\,III we give the general solution. Section IV
examines certain limiting cases, starting from the linearization 
of the problem and then turning to the limits of large and small
rings. In Sec.\,V we present a general result for the minimum
value of the velocity of propagation of the wave, and finally
we summarize our results and present our conclusions in Sec.\,VI.

\section{Formulation of the problem}

Assume that we have a ring of circumference $L$ which 
contains $N$ atoms of mass $M$. Within the mean-field approximation, 
the order parameter $\Phi(x,t)$ satisfies the equation (in the range 
$-L/2 \le x \le L/2$),
\begin{eqnarray}
  i \hbar \frac {\partial \Phi(x,t)} {\partial t} =
- \frac {\hbar^2} {2 M} \frac {\partial^2 \Phi(x,t)} {\partial x^2}
+ U_0 |\Phi(x,t)|^2 \Phi(x,t),
\label{gpe}
\end{eqnarray}
with $\int_{-L/2}^{L/2} |\Phi|^2 \, dx = N$ and $U_0 > 0$, because
of the repulsive interatomic interactions that we consider. Assuming 
that $\Phi(x,t) = \Psi(z) e^{-i \mu t/\hbar}$, where $\mu$ is the 
chemical potential and $z=x-ut$,
\begin{eqnarray}
  - i \hbar u \frac {\partial \Psi} {\partial z} =
- \frac {\hbar^2} {2 M} \frac {\partial^2 \Psi} {\partial z^2}
+ (U_0 |\Psi|^2 - \mu) \Psi.
\label{gpes}
\end{eqnarray}
Here $u$ is the velocity of propagation of the assumed traveling-wave 
solution. Setting $\Psi(z) = \sqrt{n(z)} e^{i \phi(z)}$, the superfluid
velocity $v = (\hbar/M) \phi'(z)$ is given by the continuity equation
\begin{eqnarray}
  v(z) = \frac \hbar M \phi' = u + \frac {\hbar^3 C_1} {2 M^2 n(z)}.
\label{continuity}
\end{eqnarray}
There is also an Euler equation for the density,
\begin{eqnarray}
  \frac {\hbar^2} {2 M}[(\sqrt{n})']^2 + \mu n + \frac {\hbar^6 C_1^2} 
{8 M^3 n} - \frac {U_0} 2 n^2 + \frac M 2 u^2 n = C_2,
\label{euler}
\end{eqnarray}
where $C_1$ and $C_2$ are constants of integration. 

The desired solution has a vanishing derivative at the 
points $z=0$ and $\pm L/2$ where $n = n_{\rm min}$ and 
$n = n_{\rm max}$, respectively.  This allows us to 
eliminate $C_2$ from Eq.\,(\ref{euler}) and to determine that
\begin{eqnarray}
  \frac {\hbar^6 C_1^2} {8 M^3} = n_{\rm min} n_{\rm max} 
\left[\mu - \frac 1 2 U_0 (n_{\rm min} + n_{\rm max}) + 
\frac 1 2 M u^2 \right].
\label{c1}
\end{eqnarray}
This leads to 
\begin{eqnarray}
 \frac {\hbar^2} {2 M} (n')^2 = 
2 U_0 (n_{\rm max} - n) (n - n_{\rm min}) (-n + n_0),
\label{diffeq}
\end{eqnarray}
with $n_0 = (2 \mu + M u^2)/U_0 - n_{\rm min} - n_{\rm max}$.
In the case of an infinite system, $\mu = n_{\rm max} U_0$ and 
$n_{\rm max} U_0 = M u^2$, in which case Eq.\,(\ref{diffeq}) 
then reduces to the correct and known form of the differential 
equation.

\section{The general solution}

The general solution of Eq.\,(\ref{diffeq}) which is periodic in the
interval between $-L/2$ and $L/2$ is \cite{Carr}
\begin{eqnarray}
 n(z) &=& n_{\rm min} + \frac 1 m (n_{\rm max} - n_{\rm min}) 
\left[ 1 - {\rm dn}^2 \left( \frac {2 K(m) z} L \Large{|} m \right) \right]
\nonumber \\ 
&=& n_{\rm min} + (n_{\rm max} - n_{\rm min}) 
\, {\rm sn}^2 \left( \frac {2 K(m) z} L \Large{|} m \right),
\label{solution}
\end{eqnarray}
where ${\rm dn}(x|m)$ and ${\rm sn}(x|m)$ are Jacobi elliptic 
functions and $K(m)$ is the elliptic integral of the first kind. 
Since ${\rm sn}(x=0|m) = 0$ and ${\rm sn}(x=\pm K(m)|m) = 1$, 
we see that $n(z=0) = n_{\rm min}$ and $n(z=\pm L/2) = n_{\rm max}$. 
In order for Eq.\,(\ref{solution}) to be the solution of the 
differential equation of Eq.\,(\ref{diffeq}),
\begin{eqnarray}
 K(m) = \frac 1 {2 \sqrt 2} \frac L {\xi} 
\left(\frac {1 - \lambda} m \right)^{1/2},
\label{ellint}
\end{eqnarray}
where $\xi$ is the coherence length corresponding to a density 
$n_{\rm max}$, $\hbar^2/(2 M \xi^2) = n_{\rm max} U_0$, and 
$\lambda = n_{\rm min}/n_{\rm max}$. Similarly, $\mu$ and $C_1$ 
are given as
\begin{eqnarray}
 \mu = - \frac {M u^2} 2 + \frac {U_0} 2 
\left[ \left(2 - \frac 1 m \right) n_{\rm min}
+ \left( 1 + \frac 1 m \right) n_{\rm max} \right],
\nonumber \\
\label{chemu}
\end{eqnarray}
and
\begin{eqnarray}
 \frac {\hbar^6 C_1^2} {8 M^3} = \frac {U_0} 2 n_{\rm min} n_{\rm max}
\left[ \left(1 - \frac 1 m \right) n_{\rm min} + \frac 1 m n_{\rm max} \right].
\label{c1u}
\end{eqnarray}
Since $K(m) \to \pi/2$ as $m \to 0$ and $K(m) \to \infty$ as $m \to 1$, 
$m$ ranges between 0 and 1. Clearly the solution of Eq.\,(\ref{solution})
can be immediately extended to describe $\kappa$ identical solitary waves 
on a ring of circumference $\kappa L$.   

The velocity of propagation, $u$, can then be determined by integrating
the continuity equation, Eq.\,(\ref{continuity}), over one period,
\begin{eqnarray}
  \int_{-L/2}^{L/2} v \, dz &=& \frac \hbar M [\phi(L/2) - \phi(-L/2)] 
\nonumber \\
&=& u L + \frac {\hbar^3 C_1} {2 M^2}
 \int_{-L/2}^{L/2} \frac {dz} {n(z)}.
\label{velpro}
\end{eqnarray}
Assuming that $\phi(L/2) - \phi(-L/2) = 2 \pi q$ with $q$ an
integer as a consequence of the periodic boundary conditions, 
we solve the above equation in terms of $C_1$ and combine it 
with Eq.\,(\ref{c1}) to obtain 
\begin{eqnarray}
 \frac u c = \frac {2 \sqrt 2 \pi q \xi} L 
 \pm \frac 1 L \left[ \lambda 
+ \frac {1- \lambda} m \right]^{1/2} 
\int_{-L/2}^{L/2} \frac {\sqrt{n_{\rm min} n_{\rm max}}} 
{n(z)} \, dz.
\nonumber \\
\label{velprof}
\end{eqnarray}
Here $c$ is the speed of sound of a homogeneous gas of density 
$n_{\rm max}$, $M c^2 = n_{\rm max} U_0$. We remark from 
Eq.\,(\ref{velpro}) that $C_1$ and $u$ have opposite signs when 
$q=0$.

While the parameterization of $n(z)$ that we adopt above is natural 
mathematically, it is of greater physical interest to explore solutions 
for fixed $N$, $L$, and $U_0$ as a function of, e.g., $n_{\rm max}$. To 
this end, it is convenient to pick a value of $n_{\rm max}$ and $m$, and 
determine $n_{\rm min}$ from Eq.\,(\ref{ellint}). It is clear from its 
definition that $n_{\rm min}$ must be greater than or equal to zero for 
physically meaningful solutions. The value of $m$ can then be adjusted 
to satisfy the normalization condition on $n(z)$. 

Equations (\ref{solution}), (\ref{ellint}), and (\ref{velprof})
are consistent with Bloch's theorem \cite{Bloch}. Because of the 
periodic boundary conditions imposed, the order parameter 
$\Psi_q(z)$ of the solitary wave corresponding to the branch 
with quantum number $q$ is given by that obtained for $q=0$ 
by exciting the center of mass motion, $\Psi_q(z)= e^{2 \pi i q z/L} 
\Psi_0(z)$. These two solutions have the same density, but 
there is a difference in the velocity of propagation so that 
$[u(q)-u(q=0)]/c = 2 \sqrt 2 \pi q \xi/L$, in agreement 
with Eq.\,(\ref{velprof}). Similarly, the energy spectrum consists 
of a periodic part plus an envelope function which results 
from the energy of the center of mass motion. 

\section{Limiting cases}

\subsection{Linearization of the problem}

When $n_{\rm min} \to n_{\rm max}$, one can obtain equations 
for the force and for continuity by linearizing the problem in small 
deviations of the density and the phase from the homogeneous 
solution $\Psi_0 = R_0 e^{i \phi_0}$,
\begin{eqnarray}
\Psi = (R_0 - \delta R) \,\, e^{i (\phi_0 + \delta \phi)}  
      \approx \sqrt{n_0} \, (1 - \alpha + i \delta \phi),  
\label{gpelin}
\end{eqnarray}
where $\alpha = \delta R/R_0$ and $\phi_0 = 0$.
When the two are combined, we obtain 
\begin{eqnarray}
\frac {\hbar^2} {2M} \alpha''' + 2 (Mu^2 - n_{\max} U_0) \alpha' = 0.  
\label{gpelinn}
\end{eqnarray}
Thus, $\alpha' = C \sin(k z + \phi_0)$ with $C$ and $\phi_0$ 
constants. Periodicity requires that $k L = 2 \pi l$ with 
$l$ an integer. Here, we set $l=1$ so that
\begin{eqnarray}
\left( \frac{u}{c} \right)^2 = 1 +\frac{2 \pi^2 \xi^2}{L^2}.
\label{velprolin}
\end{eqnarray}
 In the limit of ``large" rings, i.e., $L \gg \xi$, 
\begin{eqnarray}
\frac{u}{c}  \approx 1 + \frac {\pi^2 \xi^2} {L^2}.
\label{velsm}
\end{eqnarray}
In the opposite limit of ``small" rings, $L \ll \xi$, then 
\begin{eqnarray}
\frac{u}{c} \approx \sqrt{2} \pi \frac{\xi}{L} 
\left( 1 + \frac {L^2} {4 \pi^2 \xi^2} \right).
\label{velsmm}
\end{eqnarray}
We note that Eq.\,(\ref{velprolin}) can also be derived 
directly from Eq.\,(\ref{velprof}) provided that one considers a
fixed length $L$ and then takes $n_{\rm min} \to n_{\rm max}$, which
implies that $m \to 0$ in Eqs.\,(\ref{solution}) and (\ref{ellint}).

\subsection{Limit of large rings}

We now consider the limiting forms of the solution appropriate 
for large and small rings.  For large rings, $L \gg \xi$, and the 
general solution of Eq.\,(\ref{solution}) reduces to 
\begin{eqnarray}
 n(z) = n_{\rm min} + (n_{\rm max} - n_{\rm min}) 
{\rm tanh}^2 \left( \frac z \zeta \right), 
\label{profinf}
\end{eqnarray}
where $\zeta = \sqrt 2 \xi/(1 - n_{\rm min}/n_{\rm max})^{1/2}$. 
This is in agreement with the known result, plus corrections of order 
$e^{-L/\zeta}$. In this limit, $m \to 1$ with 
\begin{eqnarray}
m \approx 1 - 16 \exp{\left[ - \frac {L} {\sqrt 2 \xi} 
   \left( 1 - \lambda \right)^{1/2} \right] }
\label{m}
\end{eqnarray}
and also $K(m) \approx \ln (4 / \sqrt{1-m})$. From Eq.\,(\ref{velprof}) 
we find that
\begin{eqnarray}
 \frac u c &\approx& 2 \sqrt 2 \pi q \frac {\xi} L +
\sqrt{ \frac {n_{\rm min}} {n_{\rm max}} } +
\nonumber \\
&+& \frac {2 \sqrt 2 \xi} L \tan^{-1} \left[
\sqrt{ \frac {n_{\rm max}} {n_{\rm min}} - 1} \, \, 
{\rm tanh} \left( \frac L {2 \zeta} \right)
\right],
\label{vellargee}
\end{eqnarray}
which reduces to the familiar result $u/c = \sqrt{{n_{\rm min}}
/{n_{\rm max}}}$ in the limit of an infinite ring.  

Figure 1 shows $u/c$ as a function of $n_{\rm max}$ 
for the parameter choice $N/L=1$, $L=20$, $U_0 = 0.6$,
and $q=0$ (with $\hbar = M = 1$) so that $L/\xi > 20$.
Exact results (dots) derived from Eq.\,(\ref{velpro}), 
and values obtained using the approximation of 
Eq.\,(\ref{vellargee}) are given. The agreement is remarkably 
good except in the limit $n_{\rm max} \to n_{\rm min}$. In 
this limit, Eq.\,(\ref{vellargee}) yields $u/c = 1$, which 
does not contain the higher-order quadratic term seen in 
Eq.\,(\ref{velsm}). In the opposite limit, where $n_{\rm min} 
\to 0$, Eq.\,(\ref{vellargee}) gives the exact result, as 
shown in Sec.\,V.
 
In the limit of small $n_{\rm min}$ Eq.\,(\ref{vellargee}) 
reduces to
\begin{eqnarray}
 \frac u c \approx \frac {\sqrt 2 \pi \xi} L (1 + 2q) +
\sqrt{ \frac {n_{\rm min}} {n_{\rm max}} }.
\label{vellargeeee}
\end{eqnarray}
Consider first the choice $q=0$.  In this case
$n_{\rm min}$ can be arbitrarily small in analogy with a 
``dark" solitary wave in the $L \to \infty$ limit. The 
velocity of propagation is bounded from below by $\xi/L$, 
$u_{\rm min}/c = \sqrt 2 \xi \pi/L$, which corresponds to the
minimum value of $u/c$ shown in Fig.\,1. Equivalently,  
\begin{eqnarray}
\frac {\hbar} {M} \phi' = u_{\rm min} = \frac {\hbar} {M} \frac {\pi} L 
\label{phasel}
\end{eqnarray}
except in the immediate vicinity of the solitary wave. This result 
is easily understood. As Eqs.\,(\ref{continuity}) and (\ref{c1u}) imply, 
when $n_{\rm min} = 0$, $C_1 = 0$ and the superfluid velocity is 
independent of position and equal to the velocity of propagation $u$.  
In the limit of a dark solitary wave, the phase $\phi (z)$ varies 
linearly, except in the limited region of small density, where the 
phase develops a discontinuity of $-\pi$. In order to satisfy the 
condition $\phi(L/2) = \phi(-L/2)$, $\phi(z)$ must have the small 
slope $\approx \pi/L$ over a large length $\approx L$. Thus, as 
Eq.\,(\ref{phasel}) indicates, $u$ cannot be zero.

\begin{figure}[t]
\includegraphics[width=8.5cm,height=5.5cm]{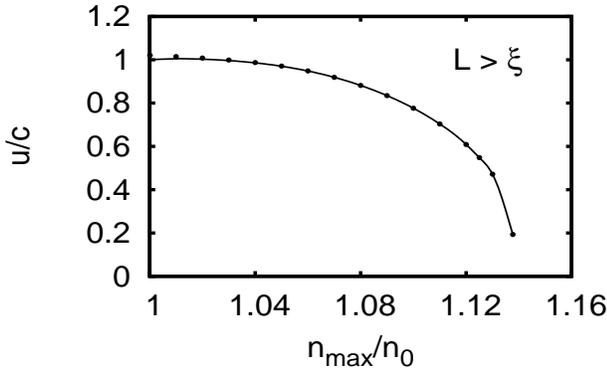}
\caption{The velocity of propagation $u/c$ in ``large" 
rings, as a function of $n_{\rm max}/n_0$, where $n_0=N/L$, 
with $n_0 = 1$, $L=20$, $U_0 = 0.6$, and $q=0$, with $\hbar 
= M = 1$. Here the dots give the exact result, Eq.\,(\ref{velpro}), 
and the solid line gives the approximate result of 
Eq.\,(\ref{vellargee}). In this figure $L/\xi > 20$.}
\label{FIG1}
\end{figure}

In the case $q \ne 0$, $u$ can vanish, but when $u$ vanishes, 
$n_{\rm min}$ is
\begin{eqnarray}
 \frac {n_{\rm min}} {n_{\rm max}} = \frac {2 \pi^2 \xi^2} {L^2} 
(1+ 2 q)^2.
\label{vellargeegenqm}
\end{eqnarray}
This result is also easy to understand. Let us set $q=-1$.  In 
this case Eq.\,(\ref{continuity}) implies that $v(z) \propto 1/n(z)$. 
Away from the center of the wave, $v(z)$ is constant and equal to 
$(\hbar/M)(\pi/L)$ since the density is constant, but the phase 
$\phi(z)$ is equal to $\pi (z/L)$. The total accumulation of the 
phase over $L$ is thus approximately $\pi$, while another $\pi$ is 
needed in order to create the total phase difference of $2 \pi$. 
This is provided from the region around the wave where $v(z) 
\approx (\hbar/M) (\pi/L) \, n_{\rm max}/n(z)$, within a length 
scale of order $\xi \sqrt {n_{\rm min}/n_{\rm max}}$. This is 
the length scale of the dominant integration interval of $v(z)$
around its singularity. Since $\phi'$ is $\approx (\pi/L) 
(n_{\rm max}/n_{\rm min})$ in this regime, and in order for the 
integral of $\phi'(z)$ to be equal to $\pi$ when we integrate 
it around a length scale of order $\xi \sqrt {n_{\rm min}/
n_{\rm max}}$, we find that  
\begin{eqnarray}
 \int \phi'(z) \, dz = \pi \approx \left( \frac {\pi} L 
\frac {n_{\rm max}} {n_{\rm min}} \right) 
\left( \xi \sqrt{\frac {n_{\rm min}} {n_{\rm max}}} \right).
\label{explanation}
\end{eqnarray}
This implies that $n_{\rm min}/n_{\max} \sim (\xi/L)^2$.

These results agree with those of Carr {\it et al.} \cite{Carr}.
In that study only static solitary waves were considered, and  
(as mentioned in Ref.\,\cite{Carr}) this is not possible for zero 
phase difference.  Their choice of $\xi/L = 1/25$ 
leads to $n_{\rm min}/n_{\rm max} \approx 0.032$ according to 
Eq.\,(\ref{vellargeegenqm}), which is in good agreement with 
the numerical results of Fig.\,4 of this reference. 

It should also be noted that Eq.\,(\ref{vellargeegenqm}) is valid
for a single solitary-wave solution. For a two-solitary wave solution
it is possible for them to be both static and dark (i.e., to have
a node in the density).

The maximum possible value of $u$ is achieved
in the limit of sound waves $n_{\rm min} \to n_{\rm max}$ where 
\begin{eqnarray}
 \frac {u_{\rm max}} c = 1 + 2 \sqrt 2 \pi q \frac \xi L 
+ \frac {\pi^2 \xi^2} {L^2},
\label{ulinnngen}
\end{eqnarray} 
in agreement with Eq.\,(\ref{velsm}). Finally, we note that 
the normalization condition takes the form $N/L = 
n_{\rm max} \left[ 1 - 2 \sqrt 2 \, ({\xi}/L) 
\left( 1 - \lambda \right)^{1/2} \right]$ in this limit.

\begin{figure}[t]
\includegraphics[width=8.5cm,height=5.5cm]{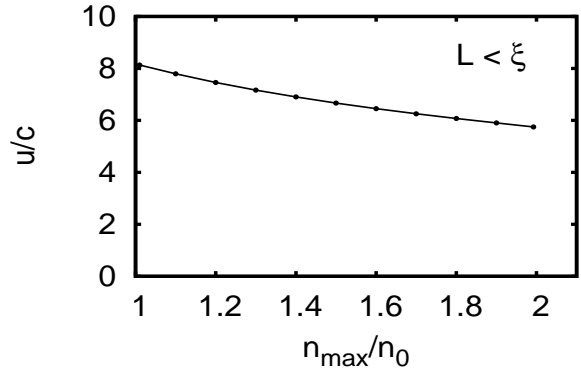}
\caption{The velocity of propagation $u/c$ in ``small" rings,
as a function of $n_{\rm max}/n_0$, where $n_0=N/L$, with 
$n_0 = 1$, $L=0.5$, $U_0 = 0.6$, and $q=0$, with $\hbar = M 
= 1$. Here the dots give the exact result, Eq.\,(\ref{velpro}), 
and the solid line gives the approximate result of 
Eq.\,(\ref{velsmalll}). In this figure $L/\xi < 0.775$.}
\label{FIG2}
\end{figure}

\subsection{Limit of small rings}

For small rings, $L \ll \xi$, the general solution of 
Eq.\,(\ref{solution}) reduces to 
\begin{eqnarray}
 n(z) = n_{\rm min} + (n_{\rm max} - n_{\rm min}) 
\sin^2 \left( \frac {\pi z} L \right),
\label{profilesmall}
\end{eqnarray}
plus corrections of order $m$ (i.e., of order $(L/\xi)^2$)
since $ m \approx (1 - \lambda) (L^2/2 \pi^2 \xi^2)$ and 
$K(m) \approx (\pi/2) (1 + m/4)$ in the limit $m \to 0$. 
Turning to the velocity of propagation of the wave, 
Eq.\,(\ref{velprof}) implies that in the limit of small
rings
\begin{eqnarray}
 \frac u c &\approx& \frac {\sqrt 2 \pi \xi} L 
( 1+ 2q ) + \frac L {2 \sqrt 2 \pi \xi} 
\sqrt{ \frac {n_{\rm min}} {n_{\rm max}}}. 
\label{velsmallll}
\end{eqnarray}
One can identify the term proportional to $1+ 2q$ as the difference 
$(E_{q+1} - E_q)/\delta p$, where $E_q = \hbar^2 q^2/(2 M R^2)$ 
is the single-particle kinetic energy of a particle in a ring of 
radius $R$ with angular momentum $\hbar q$, and $\delta p = 
\hbar/R$. This expression coincides with Eq.\,(\ref{velsmm}), 
found by linearization, when $q=0$ and $n_{\rm min} \to 
n_{\rm max}$. Here, $n_{\rm min}$ is not bounded from below as 
it was in the limit of large rings. Remarkably, $u$ saturates 
to a constant value independent of $n_{\rm min}/n_{\rm max}$ 
when $L/\xi \to 0$. This happens because 
\begin{eqnarray}
 \frac u c &\approx& 2 \sqrt 2 \pi q \frac {\xi} L + 
\frac 1 \pi  \left(\frac {1 - \lambda} m \right)^{1/2}
\int_{-\pi/2}^{\pi/2} \frac {\sqrt{\lambda} \, dw} 
{\lambda + (1-\lambda) \sin^2 w}
\nonumber \\
   &=& \frac {\sqrt 2 \pi \xi} L (1 + 2q) 
\label{velsmalll}
\end{eqnarray}
in this limit, since the integral is independent of 
$\lambda$. In other words, in the limit of small rings, the
sinusoidal dependence of the density always gives rise to a
phase difference of $\pi$ for any value of the ratio $n_{\rm min}
/n_{\rm max}$. Since the periodic boundary conditions require
that the total phase difference has to be $2 \pi q$, the additional 
phase accumulation must be $(1 + 2 q)\pi$. This fact gives 
a slope to the phase that is $\approx (1+ 2 q) \pi / L$ 
(plus corrections of order $(L/\xi)^2$, and therefore 
$u \approx (1+ 2 q) (\hbar/M) (\pi/L)$, in agreement with 
Eq.\,(\ref{velsmalll}). Finally, the normalization condition 
is $N/L = (n_{\rm min} + n_{\rm max})/2$ in this limit.

Figure 2 shows $u/c$ as a function of $n_{\rm max}$ 
for the parameter choice $N/L=1$, $L=0.5$, $U_0 = 0.6$, and $q=0$ 
(with $\hbar = M = 1$), so that $L/\xi < 0.775$. Exact results
(dots) and values obtained using the approximation of 
Eq.\,(\ref{velsmallll}) are given. The agreement is remarkably 
good in both limits $n_{\rm max} \to n_{\rm min}$, and 
$n_{\rm min} \to 0$, as Eq.\,(\ref{velsmallll}) gives the exact
result in both cases, as Eqs.\,(\ref{velsmm}) and (\ref{limitu}) 
indicate.

\section{A general result}

It is interesting to note that, in the limit of a dark 
solitary wave ($n_{\rm min} \to 0$), the velocity of propagation 
$u$ of the wave is given by the same formula, 
\begin{eqnarray}
 \frac {u_{\rm min}} c = \frac {\sqrt 2 \pi \xi} L (1 + 2 q),
\label{limitu}
\end{eqnarray}  
or
\begin{eqnarray}
 \phi' = \frac \pi L (1+ 2 q),
\label{limituu}
\end{eqnarray}  
in the limits of both small and large rings, as indicated  
by Eqs.\,(\ref{vellargeeee}) and (\ref{velsmallll}).
In fact, Eq.\,(\ref{limitu}) gives the minimum possible value 
of $u$ for {\it any} $L$: A change of variable tranforms 
Eq.\,(\ref{velprof}) into
\begin{eqnarray}
  \frac u c = \frac {2 \sqrt 2 \pi q \xi} L + 
\frac 1 {2 K(m)} \left[ \lambda + \frac {1 - \lambda} m \right]^{1/2} 
\times \nonumber \\
\int_{-1/\sqrt{\lambda}}^{1/\sqrt{\lambda}} \frac {dw} 
{[1 + (1-\lambda) w^2] \sqrt{1 - \lambda w^2} \sqrt{1 - \lambda m w^2}}.
\label{tr}
\end{eqnarray}
The integral can be evaluated in the limit $\lambda \to 0$ 
by closing the integration contour with a semicircle 
in the upper half complex plane. The only contribution 
to the integral comes from the pole at $w = i$. In this way, 
one obtains Eq.\,(\ref{limitu}) for any finite value of $L$. 

In the limit where $n_{\rm min} \to 0$, the integral in 
Eq.\,(\ref{velprof}) is dominated by the behavior of the density 
near $z = 0$ since ${\rm sn}^2(x|m) \approx x^2$ for $x \to 0$. 
As a result, the integral is approximately proportional to 
$\int_{-\infty}^{\infty} dz/(\lambda + z^2) = \pi/\sqrt{\lambda}$. 
As a result, there is a contribution to $\phi'$ which is equal to $\pi/L$.  
For a nonzero $q$, there is an additional contribution of $2 \pi q/L$
to $\phi'$, from which Eqs.\,(\ref{limitu}) and (\ref{limituu}) follow.

\section{Summary and conclusions}

To summarize, solitary-wave solutions of the nonlinear
Schr\"odinger equation framework show interesting 
features if one imposes the periodic boundary conditions 
appropriate for the description of Bose-Einstein condensed 
atoms confined in tight toroidal traps. 

In the limit of large rings, where the size of the wave is
much smaller than the length of the ring, the density is 
exponentially localized on a length scale determined 
by the coherence length and is constant over the remainder 
of the ring. Depending on the phase difference at the edges of the
period, either the minimum density can vanish (with the velocity 
of propagation being nonzero) or the velocity of propagation
can vanish (with the minimum density being nonzero).  However,  
both cannot be zero simultaneously.

In the limit of small rings, the density is sinusoidal 
within a length scale given by the circumference of the ring 
itself. The velocity of propagation saturates to a constant
value, which is independent of $n_{\rm min}$. For this reason, 
the size of the ring must exceed some minimum size in order 
to create a static wave with $u = 0$ \cite{Carr}. The minimum size 
can be determined from Eq.\,(\ref{tr}) for $q = -1$ and in the limit 
$m \to 0$ and $\lambda \to 1$ with $m/(1-\lambda) = 3$. 
Under these conditions, $u = 0$ and $L/\xi = \sqrt 6 \, \pi$, 
in agreement with Ref.\,\cite{Carr}.
  
In the limit of an infinite ring, the velocity of propagation of
the solitary wave depends on the ratio $n_{\rm min}/n_{\rm max}$.
In the opposite limit of small rings, the range of possible 
values of the velocity of propagation becomes narrower as $L$ 
decreases. In the limit of very small $L$, $u$ saturates to a 
constant value and becomes independent of $n_{\rm min}/n_{\rm max}$.

The general result that the minimum value of $u$ is inversely 
proportional to $L$ may be relatively easy to investigate 
experimentally. 

Finally, we note that as opposed to the case
of an infinite ring, which contains only the dimensionless quantity 
$n_{\rm min}/n_{\rm max}$, a second dimensionless quantity, 
$L/\xi$, appears in the present problem.  In particular, 
in the limit of sound waves (where $1 - n_{\rm max}/n_{\rm max} 
\to 0$) and large rings (where $L/\xi \to \infty$), the product 
of these two terms appears, and the two limits ``compete".  
Different answers will result depending on the way in which these 
limits are taken.  These different answers are not only of theoretical 
interest; they represent different physical situations determined by 
the way that an experiment is performed.

\acknowledgements

We thank N. Efremidis, W. von Klitzing, and S. Komineas for
useful discussions.


\begin{thebibliography}{99}

\bibitem{ZS} V. E. Zakharov and A. B. Shabat, Zh. Eksp. Teor. 
Fiz. {\bf 64}, 1627 (1973) [Sov. Phys. JETP {\bf 37}, 823 (1973)].

\bibitem{Ts} T. Tsuzuki, J. Low Temp. Phys. {\bf 4}, 441 (1971).

\bibitem{Lieb} E. Lieb, Phys. Rev. {\bf 130}, 1616 (1963).

\bibitem{Kul} P. P. Kulish, S. V. Manakov, and L. D. Faddeev, 
Theor. Math. Phy. {\bf 28}, 615 (1976).

\bibitem{Ish} M. Ishikawa and H. Takayama, J. Phys. Soc. Jpn. {\bf 49}, 
1242 (1980).

\bibitem{Fra} D. J. Frantzeskakis, J. Phys. A: Math. Theor. {\bf 43}, 
213001 (2010).

\bibitem{Han} S. Burger, K. Bongs, S. Dettmer, W. Ertmer, 
K. Sengstock, A. Sanpera, G. V. Shlyapnikov, and M. Lewenstein,
Phys. Rev. Lett. {\bf 83}, 5198 (1999)

\bibitem{Phil} J. Denschlag, J. E. Simsarian, D. L. Feder, 
C. W. Clark, L. A. Collins, J. Cubizolles, L. Deng, E. W. Hagley,  
K. Helmerson, W. P. Reinhardt, S. L. Rolston, B. I. Schneider, and
W. D. Phillips, Science {\bf 287}, 97 (2000).

\bibitem{Kurn} S. Gupta, K. W. Murch, K. L. Moore, T. P. Purdy, and 
D. M. Stamper-Kurn, Phys. Rev. Lett. {\bf 95}, 143201 (2005).

\bibitem{Olson} S. E. Olson, M. L. Terraciano, M. Bashkansky, 
and F. K. Fatemi, Phys. Rev. A {\bf 76}, 061404(R) (2007).

\bibitem{wvk} I. Lesanovsky and W. von Klitzing, Phys. Rev. Lett. 
{\bf 99}, 083001 (2007).

\bibitem{Hen} K. Henderson, C. Ryu, C. MacCormick, and M. G. Boshier,
New J. Phys. {\bf 11}, 043030 (2009).

\bibitem{Phillips} C. Ryu, M. F. Andersen, P. Clad\'e, V. Natarajan, 
K. Helmerson, and W. D. Phillips, Phys. Rev. Lett. {\bf 99}, 
260401 (2007).

\bibitem{Carr} L. D. Carr, C. W. Clark, and W. P. Reinhardt,
Phys. Rev. A {\bf 62}, 063610 (2000).

\bibitem{Bloch} F. Bloch, Phys. Rev. A {\bf 7}, 2187 (1973).

\end{thebibliography}
\end{document}